\documentclass[%
 reprint,
superscriptaddress,
 amsmath,amssymb,
 aps,
 twocolumn
]{revtex4-2}

\usepackage[]{bm}
\usepackage[]{color}
\usepackage{graphicx}

\usepackage{titlesec}

\newcommand{\Bb}{\mathbf{B}}
\newcommand{\Eb}{\mathbf{E}}
\newcommand{\kb}{\mathbf{k}}
\newcommand{\qb}{\mathbf{q}}
\newcommand{\ii}{\textrm{i}}
\newcommand{\zu}{\hat{\mathbf{z}}}

\begin{document}

\title{
    X-ray reflection: a FLUKA model and its application in the design
    of\\synchrotron light beamlines and CERN's Future Circular Collider
}

\author{G.~Mazzola}
\email{Contact author: giuseppe.mazzola@cern.ch}
\affiliation{European Organization for Nuclear Research, Esplanade des Particules 1, 1211 Geneva 23, Switzerland}
\affiliation{Facultat de F\'isica, Universitat de Barcelona, Martí i Franquès 1-11, 08028 Barcelona, Spain}

\author{S.~Chitra}
\affiliation{Argonne National Laboratory, 9700 S. Cass Avenue, 60439 Lemont, IL, USA}

\author{A.~Devienne}
\affiliation{European Organization for Nuclear Research, Esplanade des Particules 1, 1211 Geneva 23, Switzerland}

\author{A.~Frasca}
\affiliation{European Organization for Nuclear Research, Esplanade des Particules 1, 1211 Geneva 23, Switzerland}
\affiliation{University of Liverpool, Brownlow Hill, L69 7ZX Liverpool, UK}

\author{M.~J.~García-Fusté}
\affiliation{ALBA Synchrotron, Carrer de la Llum 2–26, 08290 Cerdanyola del Vallès, Spain}

\author{D.~Heinis}
\affiliation{ALBA Synchrotron, Carrer de la Llum 2–26, 08290 Cerdanyola del Vallès, Spain}

\author{A.~Lechner}
\affiliation{European Organization for Nuclear Research, Esplanade des Particules 1, 1211 Geneva 23, Switzerland}

\author{G.~Lerner}
\affiliation{European Organization for Nuclear Research, Esplanade des Particules 1, 1211 Geneva 23, Switzerland}

\author{L.~Rebuffi}
\affiliation{Argonne National Laboratory, 9700 S. Cass Avenue, 60439 Lemont, IL, USA}

\author{M.~Sanchez del Rio}
\affiliation{European Synchrotron Radiation Facility, 71 Avenue des Martyrs, 38043 Grenoble, France}

\author{D.~L.~Windt}
\affiliation{Reflective X-ray Optics LLC, 425 Riverside Dr., 10025 New York, USA}

\author{E.~Graug\'es}
\affiliation{Facultat de F\'isica, Universitat de Barcelona, Martí i Franquès 1-11, 08028 Barcelona, Spain}

\author{F.~Salvat~Pujol}
\affiliation{European Organization for Nuclear Research, Esplanade des Particules 1, 1211 Geneva 23, Switzerland}

%\date{\today}

\begin{abstract}
Relying on atomic scattering factors from evaluated databases, a
new model for the reflectivity of x~rays on solid surfaces has
been developed for FLUKA~v4-6.0. This model accounts for the
variation of reflectivity as a function of the photon energy,
its incidence angle, and linear polarisation; surface roughness
effects are also taken into account.
FLUKA reflectivities agree well with those obtained from
state-of-the-art codes used for the characterization of optical devices, both
for homogeneous solids and for multilayer mirrors. This new capability renders
FLUKA a nearly one-stop shop for synchrotron radiation
simulations: emission from bending magnets and wigglers, photon
transport and interaction, electromagnetic (and hadronic when
applicable) shower development in complex geometries, as well as
x-ray reflection at designated solid surfaces can now be all
accounted for in a single FLUKA run. This streamlined FLUKA
simulation workflow greatly simplifies the plethora of
simulation tools that Monte Carlo practitioners previously needed to
rely on. Two application scenarios of this new reflectivity model are
showcased: first, the use of a multilayer mirror to deflect
x~rays from an optical hutch onto an experimental hall at the
MINERVA beamline of the ALBA synchrotron and, second, the
assessment of the photon flux near the interaction point
at the CERN's Future Circular Collider (in its electron-positron
stage) as a result of upstream x-ray reflections.
\end{abstract}


\maketitle

\section{Introduction}
\label{sec:intro}

X-ray reflection is routinely exploited in synchrotron-radiation (SR)
light sources for beam manipulation purposes: carefully positioned and
oriented mirrors and multilayers are used to steer incoming x~rays from
an optical hutch towards experimental halls, and to perform wavelength
filtering in the case of multilayers. Particle-transport
simulations for such scenarios often involve a combination of
tools~\cite{arnaud, sunil1, sunil2}: a code for the generation of the SR
source term~\cite{stac8,xop,FLUKA1,FLUKA2,FLUKA3,Hugo2025}, a
general-purpose code for photon transport through the beamline and
optical hutch~\cite{FLUKA1,FLUKA2,FLUKA3,Hugo2025,geant,phits,penelope}
and, should x-ray reflection be considered, a further code for
evaluating the x-ray reflectivity of the employed mirrors and
multilayers~\cite{stac8,xop,xoppy,imd} before resuming the transport
simulation in a further step. Alternatively, an ad-hoc analytical
synchrotron radiation shielding code~\cite{stac8} can generate the SR
spectrum from bending magnets or insertion devices and can also
calculate the reflectivity for commonly used coated mirror. However, if
detailed source terms or reflectivity from uncommon materials or complex
configurations, as multilayer, are necessary, data from specialized
codes~\cite {xop,xoppy,imd} will need to be extracted and manipulated
offline to then carry out the desired particle-transport
analysis~\cite{sunil1, sunil2}. The need for this combination of
tools renders simulation studies rather unpractical.

The reflection of x~rays also plays a surprisingly relevant role in
high-energy electron/positron colliders. For instance, at the time of CERN's
Large Electron-Positron collider (LEP)~\cite{lep}, a machine with up to
104.5~GeV per beam, it was realised that upstream x-ray reflections
substantially contribute to the photon background in the experimental
regions~\cite{vonHoltey1998}. Therefore, this effect is worth assessing in view
of the ongoing design of CERN's Future Circular Collider in its
electron-positron mode (FCC-ee)~\cite{FCCee}.

To facilitate the laborious simulation workflow outlined in the first
paragraph, while also providing a tool to address the issue raised in
the second, a dedicated model for x-ray reflection on solid surfaces has
been developed and implemented in the general-purpose particle-transport
code FLUKA~\cite{FLUKA1,FLUKA2,FLUKA3,Hugo2025}. This work is structured
as follows. In Section~\ref{sec:model}, atomic-scattering factors from
various evaluated photon data libraries are discussed and, based on
them, the model implemented in FLUKA for x-ray reflection is presented.
Its dependence on incidence angle, photon energy, and polarisation is
highlighted, both
for coated and multilayer mirrors, with surface-roughness
effects taken into account.
In Section~\ref{sec:validation}, this new FLUKA model for x-ray
reflection in transport is benchmarked and validated against standalone
state-of-the-art codes for x-ray optics~\cite{xop,xoppy,imd}. The
practical use of this novel FLUKA capability is demonstrated in two
facility-design scenarios. In Section~\ref{sec:ALBA}, a FLUKA simulation
of the MINERVA beamline~\cite{minerva} at the ALBA synchrotron is
presented, involving a multilayer mirror to steer a flux of x~rays
incoming from an optical hutch toward an experimental chamber. In
Section~\ref{sec:FCCee}, the contribution of x-ray reflection to the
photon flux around the interaction point of the FCC-ee is assessed.
Finally, a summary and conclusions are presented in
Section~\ref{sec:conclusions}.

\section{FLUKA model for x-ray reflection}
\label{sec:model}

\subsection{Optical model}

In the context of a general-purpose Monte Carlo code like FLUKA, where
materials are assumed to be homogeneous and isotropic, coherent effects
such as x-ray diffraction in crystals are generally
disregarded~\footnote{Crystal channeling and associated phenomena can
however be treated~\cite{Hugo2025}.}.
Under these assumptions, the index of refraction $n(\qb,\omega)$ of the
material encountered by an incoming x~ray can be constructed
as~\cite{henke}
\begin{equation}
    n(\qb,\omega)
    =
    1
    -
    \frac
        {r_e\lambda^2}
        {2\pi}
    \sum_{s}
        \mathcal{N}_s
        f_s(\qb,\omega)
    ,
    \label{eq:nqw}
\end{equation}
where $\omega$ is the photon frequency, $\lambda$ is its wavelength, $r_e$ is the classical electron radius, $\mathcal{N}_s$ is the
number of atoms of species $s$ per unit volume, and the sum runs over
all atomic species present in the material composition. The (complex)
atomic scattering factors $f_s(\qb,\omega)$ describe the absorption of a
photon of frequency $\omega$ by an individual atom of species
$s$~\cite{Caticha-Ellis1974,cromer,henke} and the subsequent emission
with a wavevector transfer
$\qb$. $S$-matrix calculations~\cite{kissel1995}
suggest that for photon energies strictly below the K~absorption edge (tens of
keV for atomic numbers $Z\gtrsim30$, order of keV for $30\gtrsim Z \gtrsim10$, and tens or hundreds of eV for
$Z\lesssim10$~\cite{EPDL97,epicscullen}), the variation of the atomic scattering
factor for small $\qb$ is mild. Thus, the angle-independent
atomic-scattering-factor (AIASF) approximation is adopted in this work, wherein
$f_s(\qb,\omega)$ is taken at zero momentum transfer:
\begin{equation}
    n(\omega)
    \approx
    1
    -
    \frac
        {r_e\lambda^2}
        {2\pi}
    \sum_{s}
        \mathcal{N}_s
        f_s(\omega)
    .
    \label{eq:nw}
\end{equation}
Despite the limited interval of validity for this expression, we have used it over the entire photon energy range considered in this work (from 100~eV up to 10~MeV). As shown below, the higher the photon energy, the more grazing (\textit{i.e.}, small) the angle with respect to the surface has to be for the photon to be reflected. Thus, the small-angle approximation implied by Eq.~\eqref{eq:nw} is anyway duly respected for all photon reflection energies covered in this work.

\begin{figure}
  \centering
  \includegraphics[width=\linewidth]{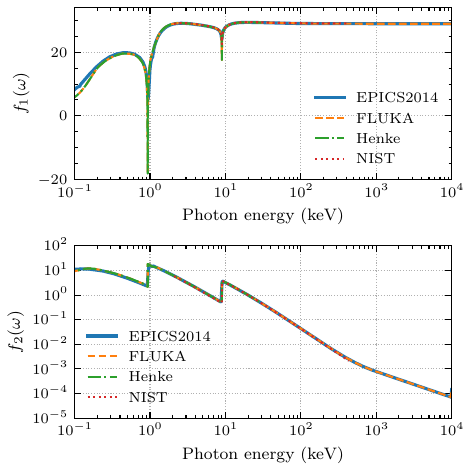}
  \caption{Real (top) and imaginary (bottom) parts of the AIASF, Eq.~\eqref{eq:henkef1f2}, for Cu. Curves represent the values from EPICS2014 (solid), the extended Henke database employed in FLUKA (dashed), the original Henke data library (dot-dashed), and NIST (dotted).}
  \label{fig:f1f2}
\end{figure}

\begin{figure}
  \centering
  \includegraphics[width=\linewidth]{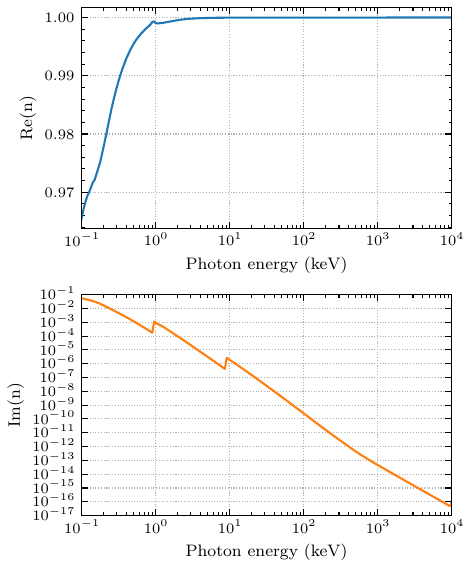}
  \caption{Real (top) and imaginary (bottom) parts of the index of
    refraction of Cu, calculated with Eq~\eqref{eq:nw} and the FLUKA
    extended Henke AIASF displayed in Fig.~\ref{fig:f1f2}.}
  \label{fig:nCu}
\end{figure}

AIASFs are complex quantities, available from state-of-the-art evaluated
atomic databases. Care should be exercised, since different libraries
encode this quantity in slightly different ways. We have adopted the
evaluated data of Henke \textit{et al.}~\cite{henke} as of August 2023~\cite{CXRO}, in which AIASFs
are tabulated as (dropping the subscript $s$ to alleviate the notation)
\begin{align}
    f(\omega)
    =
    f_1(\omega)
    +
    \ii
    f_2(\omega)
    ,
    \label{eq:henkef1f2}
\end{align}
where $\lim_{\omega\to\infty}f_1(\omega)=Z$ and
$\lim_{\omega\to\infty}f_2(\omega)=0$~\cite{kissel1995}. This library
covers photon energies from 50~eV up to 30~keV for atoms with
$1\leq Z\leq92$. We have extended the Henke library
to cover photon energies up to 10~MeV and atomic numbers up to $Z=100$,
such as to cover all material compositions allowed
in the general-purpose Monte Carlo code FLUKA. This extension,
henceforth referred to as extended Henke, has been performed relying on
the EPICS2014~\cite{epicscullen} database, wherein AIASFs are instead
defined as
\begin{align}
    f(\omega)
    =
    Z
    +
    f'(\omega)
    +
    \ii
    f''(\omega)
    ,
\end{align}
where
\begin{align}
    f'(\omega)
    =
    f_1(\omega)-Z
\end{align}
is tabulated instead of the $f_1(\omega)$ term of the Henke database,
while the adopted representation of the imaginary part remains
unaltered: $f''(\omega)=f_2(\omega)$. A Kramers-Kronig
analysis~\cite{henke,kissel1995} ensured the self-consistency of the obtained AIASF. Figure~\ref{fig:f1f2}
displays the real (top) and imaginary (bottom) parts of the AIASF
as a function of the
photon energy for Cu, in anticipation of the application discussed in
Section~\ref{sec:FCCee}. In this Figure, solid, dashed, dash-dotted and dotted
curves represent EPICS2014, extended Henke (FLUKA), the
original Henke library, and the NIST~\cite{nist} library, respectively.
Excellent agreement between EPICS2014 and Henke AIASF is observed in the range
of photon energies considered in FLUKA (above 100~eV), especially at energies
above 1~keV, thereby justifying the extension of the Henke AIASF with those of
EPICS2014 at high photon energies. The agreement extends also to the data from
the NIST library covering the energy range from 2~keV to 433~keV.
For later reference, Fig.~\ref{fig:nCu} displays the real (top) and imaginary (bottom) parts of index of refraction
of Cu, constructed with the prescription of Eq.~\eqref{eq:nw} using the
extended Henke AIASF displayed in Fig.~\ref{fig:f1f2}.

The expression laid out above to construct indices of refraction, Eq.~\eqref{eq:nw},
applies for photon energies much higher than typical binding energies
(order of eV or tens of eV). Thus, for photon energies approaching the
FLUKA photon transport limit (100~eV) one might expect binding effects
to start play a role, albeit minor. In this limit, experimental
measurements or ab-initio calculations of $n(\omega)$ should be resorted
to~\cite{adler1962}. As further stressed below, a dedicated
extension in the FLUKA implementation allows users to bypass
Eq.~\eqref{eq:nw} and load custom indices of refraction if needed.

\subsection{X-ray reflectivity on the surface to homogeneous media}

\begin{figure}
    \centering
    \includegraphics[width=0.6\linewidth]{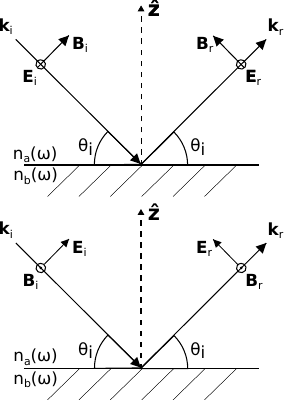}
    \caption{Schematic layout of the wavevectors and electric field
    vectors employed in the evaluation of the reflection coefficient,
    Eqs.~\eqref{eq:rsigma} and~\eqref{eq:rpi}, for
    $\sigma$- (top) and $\pi$-polarised (bottom) x rays. Magnetic field
    components shown for completeness.}
  \label{fig:sigmapi}
\end{figure}

Photon reflection on solid surfaces at an outgoing angle matching the
incoming one readily follows from the boundary conditions of the Maxwell
equations~\cite{henke,james,jackson}.
Figure~\ref{fig:sigmapi} schematically displays a photon with wavevector
$\kb_i$ propagating in a homogeneous medium~$a$ with index of refraction
$n_a(\omega)$ and impinging on a homogeneous medium~$b$, with index of
refraction $n_b(\omega)$. The incoming photon wavevector forms an incidence
angle $\theta_i$ with the planar surface. Both media are
assumed to extend sufficiently
far away from the boundary that one may disregard any potential
reflection from the other end of the media. In the context of FLUKA,
photon polarisation, albeit only linear, can be taken into account along
a user-defined direction specifying the orientation of the
electric-field vector $\Eb$, naturally normal to the photon wavevector.
The top and bottom panels of Fig.~\ref{fig:sigmapi} respectively display
an incoming polarisation state $\Eb_i$ normal ($\sigma$) and parallel
($\pi$) to the incidence plane spanned by the incoming wavevector
$\kb_i$ and the surface normal $\zu$. This Figure also displays the
wavevector $\kb_r$ of the reflected photon at an angle
$\theta_i$, along with a possible polarisation state. Magnetic field
components $\Bb_i$ are displayed for completeness. In such a scenario,
the reflection coefficients for $\sigma$ and $\pi$ polarisation follow
from the Fresnel equations~\cite{jackson}, expressed here in terms of
angles with respect to the surface (and not the surface normal):
\begin{align}
    \nonumber
    r_{\sigma}(\omega,\theta)
    &=
    \frac{E_{r\sigma}}{E_{i\sigma}}
    \\
    \label{eq:rsigma}
    &=
    \frac
        {n_a(\omega) \textrm{sin}(\theta_i) - \sqrt{n_b^2(\omega) - n_a^2(\omega)\textrm{cos}^2(\theta_i)}}
        {n_a(\omega) \textrm{sin}(\theta_i) + \sqrt{n_b^2(\omega) - n_a^2(\omega)\textrm{cos}^2(\theta_i)}},
    \\
    \nonumber
    r_{\pi}(\omega,\theta)
    &=
    \frac{E_{r\pi}}{E_{i\pi}}
    \\
    &=
    \frac
        {n_b^2(\omega) \textrm{sin}(\theta_i) - n_a(\omega)\sqrt{n_b^2(\omega) - n_a^2(\omega)\textrm{cos}^2(\theta_i)}}
        {n_b^2(\omega) \textrm{sin}(\theta_i) + n_a(\omega)\sqrt{n_b^2(\omega) - n_a^2(\omega)\textrm{cos}^2(\theta_i)}},
    \label{eq:rpi}
\end{align}
that is as the ratio of reflected to incident electric field components
for $\sigma$ and $\pi$ polarisation, respectively. From this definition, the reflected-photon electric field components follow readily:
\begin{align}
    \nonumber E_{r\sigma} &= r_\sigma E_{i\sigma},\\
 			   E_{r\pi}    &= r_\pi E_{i\pi}.
\end{align}

\begin{figure}
  \centering
  \includegraphics[width=\linewidth]{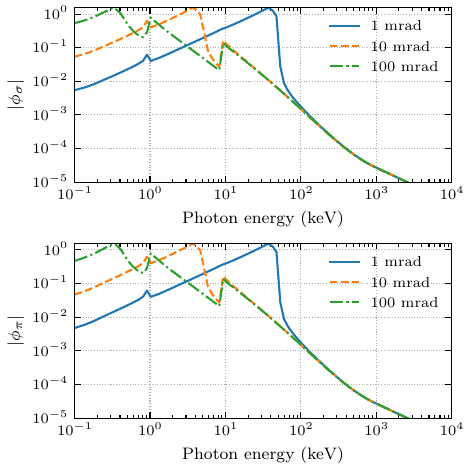}
  \caption{Absolute value of the phase shift for $\sigma$- (top) and
    $\pi$-polarised (bottom) photons reflected from a homogeneous Cu
    slab as a function of the photon energy for 1~mrad (solid), 10~mrad
    (dashed), and 100~mrad (dot-dashed) incidence angle.}
  \label{fig:phase}
\end{figure}

Incidentally, the use of a complex index of refraction, as per
Eq.~\eqref{eq:nw}, renders these expressions applicable
for both dielectric and conducting media, and yields complex reflection coefficients. Thus, a
phase shift between incident and reflected polarisation vector may
easily be incurred. Figure~\ref{fig:phase} displays the phase shift for
$\sigma$ (top) and $\pi$ (bottom) polarisation obtained as a function of
photon energy for angles ranging from 1~to 100~mrad. The phase shift
itself can even reach $\pi$ at different photon energies depending on the incidence angle. This
phase shift is disregarded here, since FLUKA currently tracks
linear photon polarisation, defined by a unit vector with real (not
complex) components. As shown below, the disregard of these
phase shifts is often inconsequent for the likelihood of reflection: it
is only problematic at very low photon energies (usually irrelevant in
practical applications) and around the Brewster angle~\cite{jackson}.

\begin{figure*}
  \centering
  \includegraphics[width=0.9\linewidth]{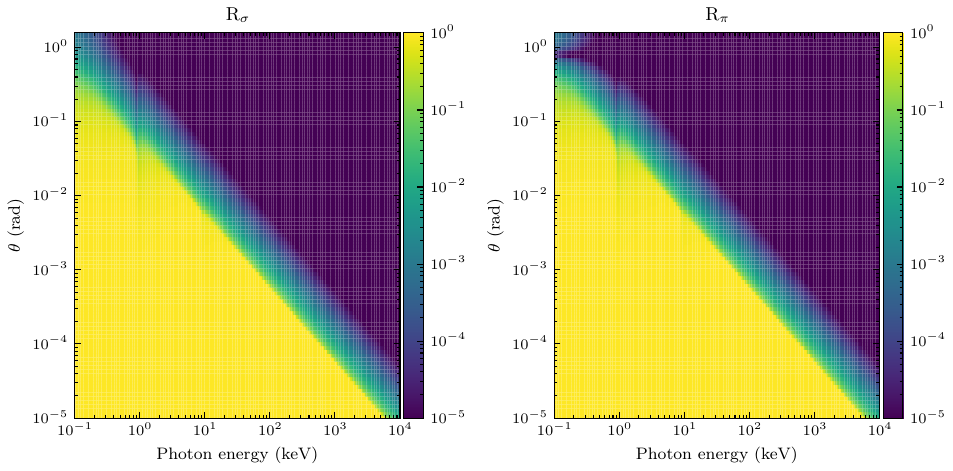}
  \caption{Photon reflectivity of Cu as a function of energy and incidence
    angle for $\sigma$ (left) and $\pi$ (right) polarisation.}
  \label{fig:R}
\end{figure*}

\begin{figure}
  \centering
  \includegraphics[width=\linewidth]{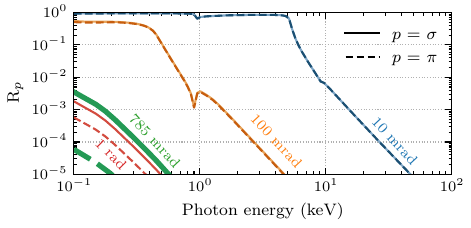}
  \caption{Photon reflectivity of Cu as a function of energy
    for $\sigma$ (solid) and $\pi$ (dashed) polarisation, for selected angles
    near the Brewster angle (thick curves).}
  \label{fig:R_1D}
\end{figure}

Not all photons impinging on a planar boundary
reflect. The reflectivity (\textit{i.e.} the likelihood of
reflection) for $\sigma$ and $\pi$ polarisation ($R_{\sigma}$ and $R_{\pi}$
respectively) follows from the reflection coefficients~\cite{jackson}:
\begin{align}
	\nonumber
    R_{\sigma}(\omega,\theta)
    &=
    |r_\sigma(\omega,\theta)|^2,
    \\
    \label{eq:Rpi}
    R_{\pi}(\omega,\theta)
    &=
    |r_\pi(\omega,\theta)|^2
    .
\end{align}
Let $P_{\sigma}$ and $P_{\pi}$ be the $\sigma$ and $\pi$
polarisation-vector amplitudes for a photon with arbitrary linear
polarisation, normalised such that $P_{\sigma}^2+P_{\pi}^2=1$. Its
reflectivity follows from Eq.~\eqref{eq:Rpi}:
\begin{equation}
    \label{eq:R}
    R(\omega,\theta) = P^2_{\sigma} R_\sigma(\omega,\theta) + P^2_{\pi} R_\pi(\omega,\theta)
    .
\end{equation}
Finally, the reflectivity for an unpolarised photon beam reads
\begin{equation}\label{eq:Runpol}
    R(\omega,\theta) = \frac{R_\sigma(\omega,\theta) +
    R_\pi(\omega,\theta)}{2}
    .
\end{equation}

Figure~\ref{fig:R} displays a bird's eye view of the $\sigma$ (left) and
$\pi$ (right) reflectivity as a function of the photon energy and
incidence angle, Eq.~\eqref{eq:Rpi},
for photons impinging from vacuum on Cu, computed using the index of
refraction shown in Fig.~\ref{fig:nCu}. The reflectivities of both polarisation states
are quite similar, except for a narrow band of angles near the Brewster
angle (785~mrad) and photon energies below 1~keV in this
example. To further elucidate this aspect, Fig.~\ref{fig:R_1D} displays the
$\sigma$ (solid curves) and $\pi$ (dashed curves) reflectivities as a
function of the photon energy for incidence angles of 10~mrad, 100~mrad,
785~mrad, and 1~rad. Indeed, accentuated discrepancies in reflectivity
are found only near the Brewster angle at photon energies well below
1~keV. Thus, since $\sigma$ and $\pi$ reflectivities coincide in most of
the energy and angular domain where the reflectivity is sizeable, the
disregard of phase shifts between the $\sigma$ and $\pi$ components
eluded to above is indeed inconsequential within the photon transport
domain of FLUKA (energies above 100~eV).  Note, however,
that the model presented here will still yield accurate reflection
coefficients as a function of linear photon polarisation (at least for
the first - and often most relevant - reflection) for photon energies
down to 100~eV.

\subsection{X-ray reflectivity on multilayer mirrors}

The model outlined above describes x-ray reflectivity on the surface of
a homogeneous solid. It has been shown that reflectivity is highest for
grazing angles and low photon energies. Multilayer mirrors (MLMs),
instead, generally allow x~rays to be reflected at larger angles and higher
photon energies~\cite{MacranderHuang2017}. In addition, MLMs exhibit x-ray wavelength
filtering capabilities and are therefore widely employed in optical
beamlines of synchrotron-light sources to select a narrow
range of photon wavelengths from the continuum SR spectrum generated by
a bending electron beam.
We have therefore extended our model to take into account x-ray
reflectivity in MLMs.

The individual building block of a MLM is a group of two (or more)
layers, typically alternating low- and high-$Z$ materials, each with a
well defined thickness, \textit{e.g.}~a Si layer of thickness $d_1$ and
a~W layer of thickness $d_2$. The stacking of $N_\ell$ such groups of
layers constitutes a MLM, typically limited by vacuum (or air) on one side
and a substrate material on the other. Thus, the number of physical
interfaces between different materials is (for two layers in the
individual building block) $2N_\ell+1$, including the interface between vacuum/air and the first layer, as well as the interface between the
last layer and the substrate.

Because of the periodic structure of a MLM, the reflectivity of an
incident photon must account for the contribution of electromagnetic
waves reflected on its various inner interfaces. Let $r_{p,2N_\ell+1}$ be the reflection coefficient of the interface between the last
layer and the substrate, $r_{p,2N_\ell}$ that of the interface between the
next-to-last and the last layer, and so on, for polarisation
$p=\{\sigma,\pi\}$. The MLM is assumed to be thin enough that photon
absorption within its volume is neglected and, furthermore, the
substrate is assumed to be effectively a semi-infinite material, so that
no photons are reflected back from its deeper regions, \textit{i.e.},
\begin{align}
    r_{p,2N_\ell+2} = 0
\end{align}
for both polarisation states $p=\{\sigma,\pi\}$. This boundary condition
allows us to evaluate the reflection coefficients of the subsequent
upper layers of the MLM. Following Parratt~\cite{parratt}, one may
recursively evaluate
\begin{align}
  r_{p,j}
    &= a_j^4 \frac{r_{p,j+1} + F_{p,j}}{1 + r_{p,j+1}F_{p,j}},
  \label{eq:rsigma_ML}
\end{align}
for $j=2N_\ell+1,2N_\ell,2N_\ell-1,...,1$, where
\begin{align}
    F_{\sigma,j}
    &= \frac{f_j - f_{j+1}}{f_j + f_{j+1}}, \\[1ex]
    F_{\pi,j}
    &=
      \frac
          {
              \displaystyle\frac{f_j}{n_j^2} - \frac{\displaystyle f_{j+1}}{n_{j+1}^2}
          }
          {
              \displaystyle \frac{f_j}{n_j^2} + \frac{\displaystyle f_{j+1}}{n_{j+1}^2}
          },\\
    f_j
    &
    = \sqrt{n_j^2 - \textrm{cos}(\theta_i)^2}
    ,
\end{align}
and $a_j = \textrm{exp}(-i \pi f_j d_j \lambda^{-1})$ accounts for the
phase shift accumulated by the photon as it traverses (half of) the
layer $j$. All $\omega$ dependencies have been temporarily dropped to
alleviate the notation. The reflection coefficients above allow us to
directly set the polarisation state of the reflected photon. Instead, the
reflectivity (\textit{i.e.}~the likelihood of reflection) of the MLM for
$\sigma$ and $\pi$ polarisation is given by
\begin{align}
    \nonumber
    R_{\sigma}&=|r_{\sigma,1}|^2,\\
    R_{\pi}   &=|r_{\pi,1}|^2
    .
\end{align}
For a photon with arbitrary linear polarisation, the reflectivity
follows from a weighted sum of the reflectivities above, as in the case
of reflection on the boundary to a homogeneous material,
Eq.~\eqref{eq:R}.

\begin{figure}
  \centering
  \includegraphics[width=\linewidth]{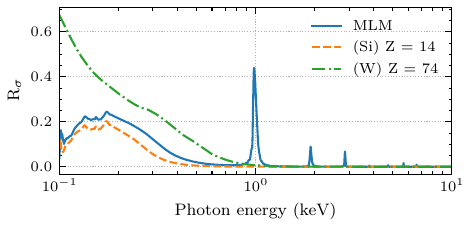}
  \caption{Reflectivity for a $\sigma$-polarised photon, incident at
    122~mrad on a multilayer mirror composed of Si-W layers (see text) on a
    Si substrate (solid line). Reflectivity of
    Si (dashed line) and W (dash-dotted line) shown for comparison.}
  \label{fig:R_ML}
\end{figure}

Figure~\ref{fig:R_ML} illustrates the reflectivity of a MLM consisting of 30~Si-W bilayers, with respective thicknesses of
4.1~nm and 1.2~nm, deposited on a Si substrate, for photon energies from
100~eV to 10~keV at an incidence angle of 122~mrad. The multilayer effects
are visible with the appearance of a conspicuous maximum around 1~keV.
Note that x-ray reflection on the boundary to a homogeneous material at
these energies would not be possible for such a large incidence angle, as
elucidated by the dashed and dot-dashed lines of the Figure,
corresponding to the reflectivity on the surface to a homogeneous slab of
Si and W, respectively.

\begin{figure*}
  \centering
  \includegraphics[width=0.45\linewidth]{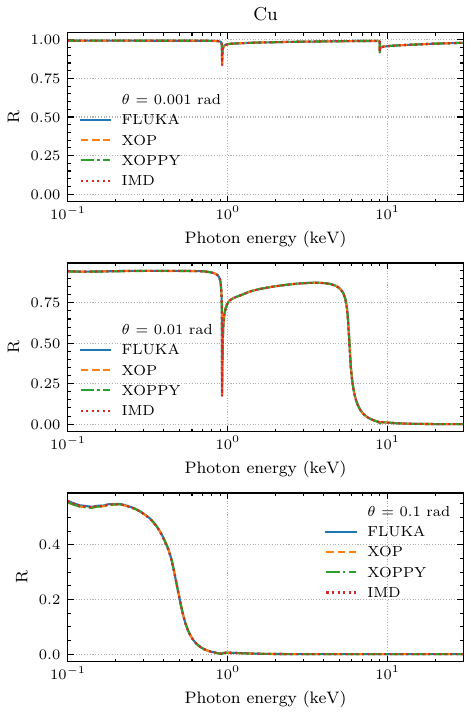}
  \centering
  \includegraphics[width=0.45\linewidth]{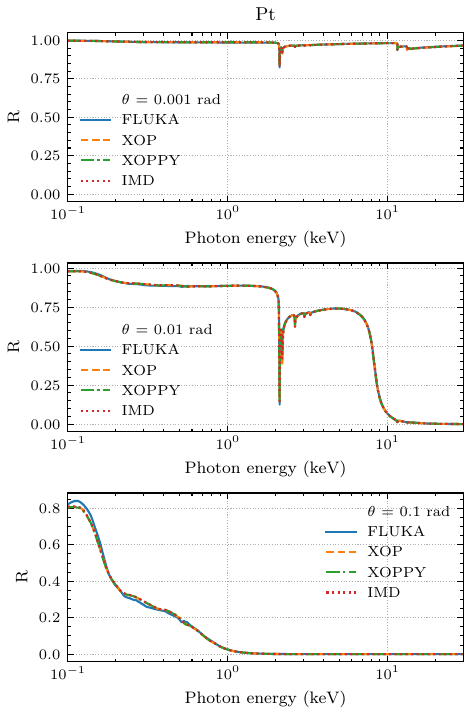}
  \caption{Reflectivity of homogeneous Cu (left) and Pt (right) slabs as
	a function of photon energy, evaluated with FLUKA (solid curves),
	XOP (dashed curves), XOPPY (dot-dashed curves), and IMD (dotted
	curves) for unpolarised photons at the indicated incidence angles.}
  \label{fig:val_R}
\end{figure*}

\begin{figure*}
    \centering
    \includegraphics[width=0.45\linewidth]{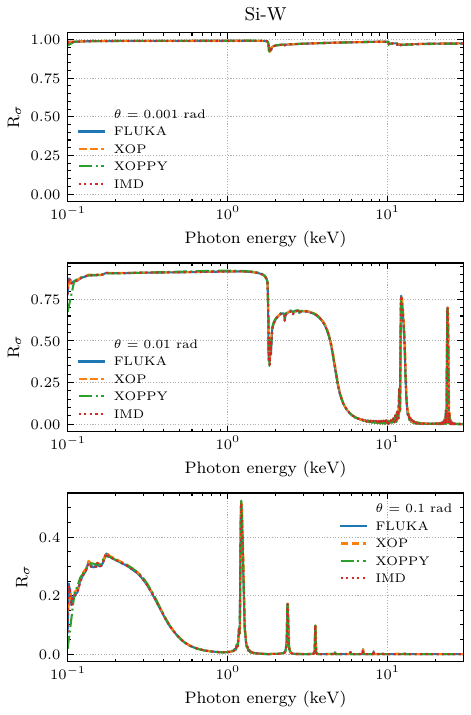}
    \centering
    \includegraphics[width=0.45\linewidth]{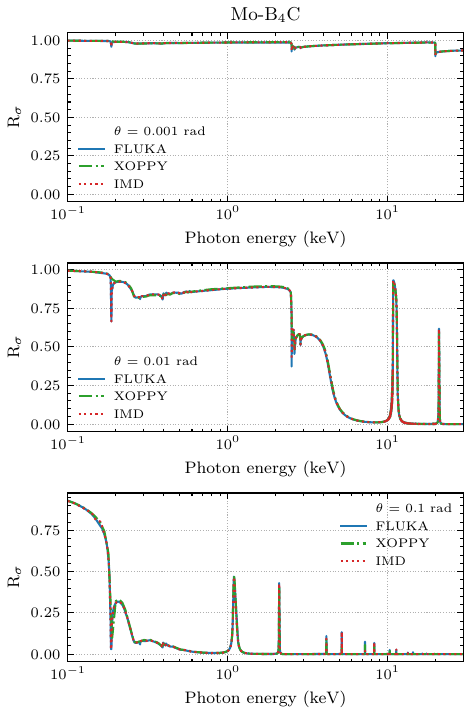}
    \caption{Reflectivity as a function of photon energy for a Si-W
    (left column) and Mo-B$_4$C (right column) multilayer mirror (see
    text) evaluated with FLUKA (solid curves), XOP (dashed curves),
    XOPPY (dot-dashed curves), and IMD (dotted curves) for
    $\sigma$-polarised photons at the indicated incidence angles.}
    \label{fig:val_ML}
\end{figure*}

\begin{figure}
    \centering
    \includegraphics[width=\linewidth]{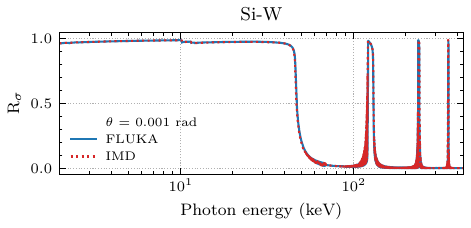}
	\caption{Reflectivity as a function of photon energy for a Si-W
	multilayer mirror (see text) evaluated with FLUKA (solid curves) and
	IMD (dotted curves) for $\sigma$-polarised photons at the indicated
	incidence angle. The IMD calculation employs NIST optical data.
	}
    \label{fig:val_ML_nist}
\end{figure}

\subsection{Surface roughness}

Finally, it should be noted that surface roughness
greatly influences the reflection coefficient,
both on homogeneous media and on MLMs. Due to this surface property, the normal
vector is no longer constant: its orientation depends on the
position on the surface. We do not expect to have a
local map of the surface roughness; instead, we adopt the effective
approach described by Nevot-Croce~\cite{Nevot} and Sinha~\cite{Sinha}, as
customary in x-ray software, whereby the reflection
coefficient is attenuated by means of a factor
\begin{equation}
  \label{eq:fNC}
    f_{NC} = \textrm{exp}(-2k_{i\perp}k_{t\perp}R_q^2),
\end{equation}
where $k_{i\perp}$ and $k_{t\perp}$ are the wavevector components normal
($\perp$) to the
reflecting surface for both incident ($i$) and transmitted ($t$) photon,
respectively, and $R_q$ is the root-mean-square (RMS) average of the
profile deviation from the mean height, as per
ISO~4287~\cite{ISO4287_1997}. The RMS average usually depends on the
manufacturing process of the specific component: it typically ranges
from 25~nm to 50~$\mu$m, although it can drop to values in the order of
a few Ångström in x-ray mirrors. Since the wavevectors can be complex
variables, the Nevot-Croce factor can itself
be a complex quantity. This implies a change for both the amplitude and
phase of the reflection coefficient. In the current FLUKA
implementation, the phase shift induced by the roughness coefficient is
disregarded for the same reasons invoked in the discussion of the
reflectivity itself. Finally, we note that the Nevot-Croce factor is
applied at every inner interface in the MLM reflectivity calculation.

\section{Model validation}
\label{sec:validation}

The x-ray reflectivity computed with the FLUKA model outlined above has
been benchmarked against state-of-the-art codes for the characterization
of optical devices, including XOP~\cite{xop}, XOPPY~\cite{xoppy}, and
IMD~\cite{imd}. These codes allow users the possibility
of selecting different optical data libraries or
even to provide their own.
To assess the differences in reflectivity
arising purely from the use of different codes, common optical data
have been used among them, namely the Henke library, covering photon
energies from 50~eV to 30~keV.
Figure~\ref{fig:val_R} displays the reflectivity of
x~rays impinging on a homogeneous slab of Cu (left) and Pt (right) as a
function of their energy, from 100~eV (lower transport limit of
photons in FLUKA) to 30~keV (upper energy limit of the Henke
library), for incidence angles from $10^{-3}$~rad (top) to
$10^ {-1}$~rad (bottom), computed with FLUKA (solid curves),
XOP~\cite{xop} (dashed), XOPPY~\cite{xoppy} (dot-dashed), and
IMD~\cite{imd} (dotted). Overall, very good agreement is observed among
the four evaluations; the residual differences in the lower right-hand panel
are due to the fact that FLUKA uses the latest version of the Henke
library (as of August 2023) while the other codes rely on an earlier
version.

Similarly, Fig.~\ref{fig:val_ML} shows the x-ray reflectivity on the
same photon energy scale and incidence angles for two MLMs: on the left
for 30~Si-W bilayers (respectively 4.1~nm and 1.2~nm thick) on a Si substrate and,
on the right, for 200~Mo-B$_4$C bilayers (respectively 2~nm and 4~nm thick) on a Si
substrate. Note the absence of XOP curves for Mo-B$_4$C since the code does
not support reflectivity calculations for
MLMs with compound materials.
The agreement among curves displayed in this Figure is also excellent,
even in the region with a rich structure of peaks.
Figure~\ref{fig:val_ML_nist} extends the
upper-left-hand panel of Fig.~\ref{fig:val_ML} to 433~keV showing curves
calculated with FLUKA and IMD. While FLUKA still relies on the extended Henke data library, IMD curve is obtained employing NIST optical
data~\cite{nist} which covers the energy range from
2~keV up to 433~keV. Note the excellent agreement between the two curves
also in this extended energy, in view of the fact that virtually
identical optical data have been used strictly within the energy
domain they cover.

While the FLUKA x-ray reflectivity model provided here is effective and
generally applicable, two extension capabilities are provided. On the
one hand, the user routine \texttt{USRCIR.f} allows
users to input their own complex index of refraction, should they have
experimental data or ab-initio calculations refining the incoherent
addition performed in Eq.~\eqref{eq:nqw}. On the other hand, the
user routine \texttt{USRREF.f} allows users to adopt
their own evaluation of the x-ray reflectivity, or to link to external
tools such as XOP, XOPPY, and IMD.

\section{Application: the MINERVA beamline at the ALBA synchrotron}
\label{sec:ALBA}

X-ray reflection at material interfaces is often
exploited in the experimental beamlines of synchrotron light source
accelerators. One or more mirrors are typically placed in the optical
hutch of the beamline to filter and select a narrow interval of photon
energies from the SR spectrum: when synchrotron
radiation photons impinge on a MLM, the
multilayer structure leads to reflection of photons within a narrow
energy band. These are then transported to the
experimental hutch in a stainless steel vacuum chamber, while any remaining
part of the spectrum must be absorbed in appropriately placed shielding.

\begin{figure}
  \centering
  \includegraphics[width=\linewidth]{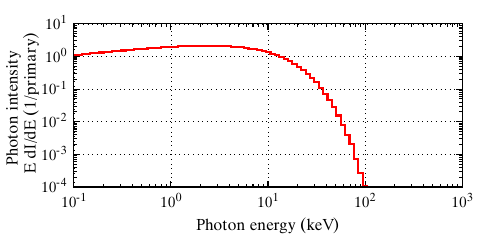}
  \caption{Synchrotron radiation spectrum generated in FLUKA during the tracking along the full trajectory of a 3~GeV electron beam in a bending magnet of the ALBA storage ring.}
  \label{fig:ALBA_SR}
\end{figure}

At present, various software tools are required to perform radiation
studies for synchrotron light beamlines involving x-ray
reflection~\cite{arnaud, sunil1, sunil2}: a code such as FLUKA,  STAC8~\cite{stac8} or
XOP~\cite{xop} for the generation of the SR spectrum
from bending magnets and wigglers (in addition to undulators
for the last two codes); a
Monte Carlo code such as FLUKA for the transport along the beamline up to the mirror;
a further code for evaluating the x-ray reflectivity, such as
XOP~\cite{xop}, XOPPY~\cite{xoppy}, or IMD~\cite{imd}; a second-step Monte Carlo simulation to transport
photons towards the experimental hall; and the necessary
scripting to coordinate data transfer among tools.

\begin{figure}
  \centering
  \includegraphics[width=\linewidth]{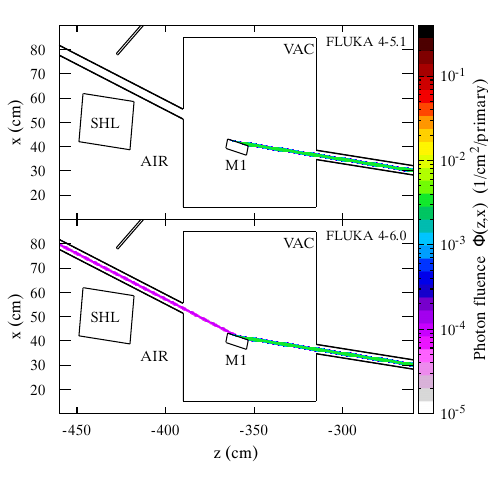}
  \caption{Photon fluence in the optical hutch of MINERVA beamline at the ALBA
    synchrotron obtained with FLUKA 4-5.1 (top) and
    FLUKA 4-6.0 (bottom).}
  \label{fig:ALBA_R}
\end{figure}

To demonstrate FLUKA's integrated capabilities for the design of SR
beamline facilities, a simplified geometry for the soft-x-ray MINERVA
beamline at the ALBA synchrotron~\cite{minerva} has been implemented in FLUKA.
The 3~GeV
electron beam was used as source term (all simulation results
below are given per primary electron). Electrons traverse a vacuum
region of 1.384~m with a magnetic field of 1.42~T normal
to the orbit plane, thereby acquiring a deflection of 11.25$^\circ$~\cite{WEPLS080}. While
traversing this effective bending magnet, electrons radiate the SR spectrum
displayed in Fig.~\ref{fig:ALBA_SR}. The emitted SR photons travel 16.66~m along
the MINERVA beamline located in port 25 of ALBA storage ring with a
restricted front end aperture of 0.48~mrad and, finally, reach the
optical hutch of the experiment. There, the photons impinge upon a
MLM (M1), consisting of 30~Si-W bilayers deposited on a Si substrate with
thickness respectively of 4.1~nm and 1.2~nm, at a grazing angle of 7~deg, as
shown in Fig.~\ref{fig:ALBA_R}.
The top panel of this Figure displays the photon fluence obtained with
FLUKA~v4-5.1, prior to the inclusion of the x-ray reflection model
presented here. In absence of x-ray reflection, the SR
photons impinging upon M1 are absorbed.
The bottom panel of Fig.~\ref{fig:ALBA_R} displays the photon fluence
obtained instead with FLUKA~v4-6.0, equipped with the x-ray reflectivity
model presented here: incoming SR photons impinging on M1 are clearly reflected
towards the entrance of the experimental hutch. Naturally, not all photons are
reflected, in accordance with the MLM reflectivity displayed in
Fig.~\ref{fig:R_ML}. The sharp peak at 1~keV in the
multilayer reflectivity allows a flux of
predominantly 1~keV photons to be reflected and transported further
along the beamline. Note that
in the geometry used for this simulation M1 is modelled considering the substrate
material (Si) exclusively: the multi-layer structure is 1~$\mu$m thin, which is
negligible as far as photon absorption and interactions are concerned.

A single FLUKA run sufficed to generate the SR spectrum from a
bending magnet, to transport SR photon along the beamline, to
handle the reflectivity on a multilayer mirror, and to further transport reflected
photons to the experimental hutch. This integrated workflow considerably simplifies
contemporary approaches relying on a concatenation of various
tools~\cite{arnaud, sunil1, sunil2}. Thus, FLUKA can be used as a single
simulation tool to track SR photons from production to the optical beamline
elements, thereby facilitating simulation studies, \textit{e.g.} the analysis of
thermal load on optical surfaces.

\section{Application: the machine-detector interface for CERN's Future
	Circular Collider (FCC-ee) project}
\label{sec:FCCee}

\begin{figure*}
  \centering
  \includegraphics[width=\linewidth]{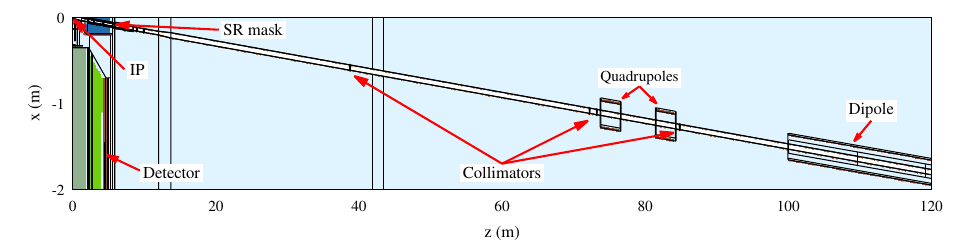}
  \caption{FLUKA geometry model of the last 120~m of the FCC-ee
    MDI region.}
  \label{fig:FCC_MDI}
\end{figure*}

At the time of CERN's Large Electron-Positron (LEP) collider~\cite{lep},
it was already noticed that SR photons, particularly in the x-ray
domain, can undergo reflection on the inner surface of the accelerator
beam pipe, allowing them to reach downstream regions far beyond the
first reflection point~\cite{vonHoltey1998}. This phenomenon, in turn,
can lead to a higher photon flux throughout the machine, an undesired
background around the interaction point (IP), and radiation damage to
the detector and the nearby beamline elements.

In view of the ongoing design of CERN's Future Circular Collider in its
electron-positron (FCC-ee) configuration, it becomes relevant to assess
the contribution of upstream x-ray reflections to the radiation
background around the IP. To address this potential issue, the geometry
of the machine-detector-interface (MDI) region of the FCC-ee was modelled
in FLUKA~\cite{frasca}, covering the last 720~m of the electron beam
line towards the IP. This region includes magnetic elements as dipoles,
quadrupoles, and sextupoles along the beamline, as well as a solenoid
field in the detector area around the IP. In addition, collimators and
SR masks are present. The FLUKA region corresponding to the inside of
the beam pipe is assigned a vacuum material; inner-beam-pipe regions
corresponding to the position of the aforementioned magnets have been
assigned a magnetic field following version 24.4 of the Global Hybrid
Correction (GHC) optical
lattice~\cite{PhysRevAccelBeams.19.111005,oidev244,CERNopt}.
Figure~\ref{fig:FCC_MDI} displays the adopted FLUKA geometry for the
last 120~m of the FCC-ee electron line towards the IP.

\begin{figure}
  \centering
  \includegraphics[width=\linewidth]{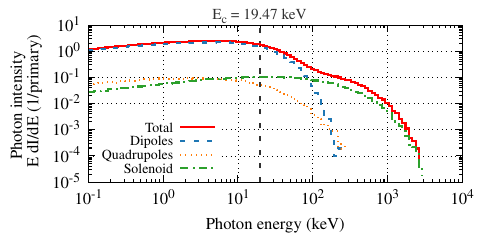}
  \caption{FLUKA SR spectrum (solid curve) generated by the passage of a
    45.6~GeV electron beam through the various magnets along the considered FCC-ee beamline. The dashed, dotted, and
    dot-dashed curve represent the contribution from dipole, quadrupole,
    and solenoid magnetic fields. The vertical line indicates the
    critical energy ($E_c$) for SR emission in the dipoles.}
  \label{fig:FCC_sr}
\end{figure}

Considering the Z-mode operation of the FCC-ee~\cite{fccee_paper}, a
45.6~GeV electron beam was used as source term in the simulations
following the aforementioned optics. SR emission in
transport~\cite{FLUKA1} was enabled in FLUKA down to a threshold energy
of 100~eV in all vacuum regions with magnetic fields. The solid
curve in Fig.~\ref{fig:FCC_sr} shows the logarithmic spectrum of
produced SR photons along the various magnets in the beamline. Non-solid curves instead display the
contributions from various magnetic field regions, namely from dipoles (dashed),
from quadrupoles (dotted), and from the solenoid field around the IP
(dash-dotted); sextupoles did not contribute appreciably. The dipole
contribution, with its distinct critical energy ($E_c$) of 19.47~keV~\cite
{Humann:2022ueh,jackson} is dominant. However, a much higher-energy contribution due
to SR emission in the solenoid field is also present, albeit of reduced
relative importance.

The fluence of photons, differential in energy and in spatial
coordinates was scored along the beamline, including the region around
the IP. Two FLUKA simulations were performed: on the one hand, a
simulation
with x-ray reflection switched off, representing the performances of
FLUKA~v4-5.1; on the other hand, a simulation
with x-ray reflection activated at all interfaces between the
beam-pipe vacuum and the materials of the beam pipe itself (Cu), as well
as collimators (Mo-C alloy), and masks (W-Ni-Cu alloy), showcasing the
enhanced capabilities of FLUKA~v4-6.0.

\begin{figure*}
  \centering
  \includegraphics[width=\linewidth]{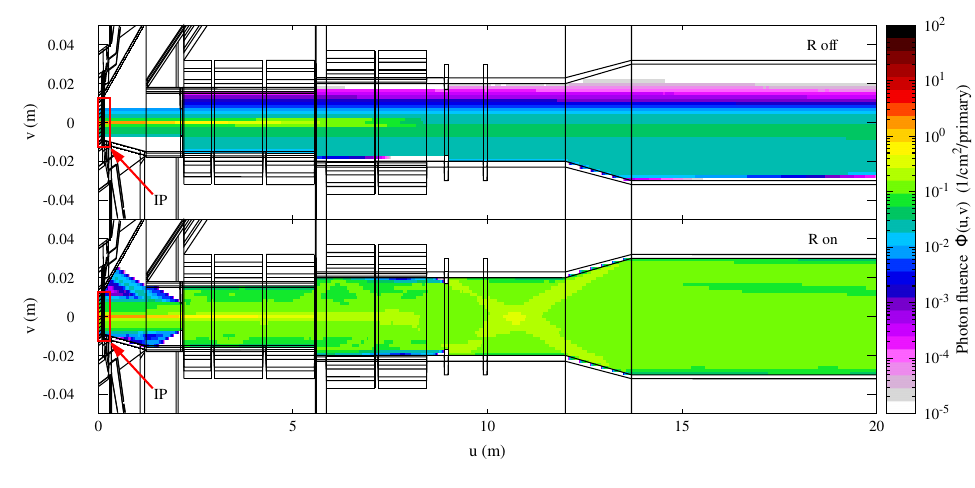}
  \caption{Spatial photon fluence in the last 20~m
    of the MDI region of the FCC-ee in Z mode, obtained in FLUKA without
    (top) and with (bottom) x-ray reflection at the different
    vacuum-material interfaces (see text).}
  \label{fig:FCC_flu}
\end{figure*}

Figure~\ref{fig:FCC_flu} displays the spatial fluence of photons in the
last 20~m leading up to the IP, with photon reflection deactivated (top
panel) and activated (bottom panel). A change of variables was performed
to transform from the slanted beamline of Fig.~\ref{fig:FCC_MDI}
(0.015~rad with respect to the $z$~axis) to the arrangement of
Fig.~\ref{fig:FCC_flu}, wherein the beamline is parallel to the $u$
axis and perpendicular to the $v$ axis. Indeed, in both panels the
photon fluence increases in the last $\sim8$~m to the IP in view of
SR production in the final-focus quadrupole magnets and in the
solenoid field around the IP, as discussed above. While the top
panel (x-ray reflection off) exhibits a fairly monotonic photon
fluence along the beamline, the bottom panel (x-ray reflection on)
shows that SR produced in the aforementioned magnetic field regions
can indeed reflect multiple times along the beam pipe. This effect
allows photons to reach downstream regions and,
specifically, the region around the IP. Furthermore, discrepancies
in photon spatial fluence due to x-ray reflectivity naturally
accentuate at off-axis positions.

\begin{figure}
  \centering
  \includegraphics[width=\linewidth]{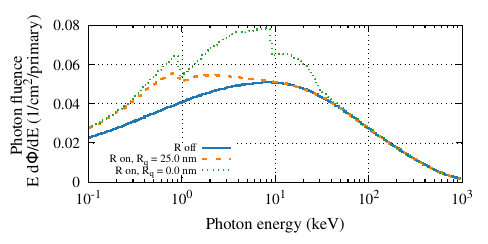}
  \caption{Energy-differential photon fluence in a cylindrical volume
    around the Interaction Point (IP) of
    FCC-ee in Z mode (see text) with x-ray reflection
    deactivated (solid line) and activated on mirror-like surfaces
    (dotted line) and with a roughness characterised by $R_q=25$~nm
    (dashed line).}
  \label{fig:FCC_sp}
\end{figure}

Figure~\ref{fig:FCC_sp} displays the energy-differential fluence of
photons in a cylindrical region surrounding the IP, with a radius $r$ of
1~cm and a length of 31~cm. This region is highlighted by a red box in
Fig.~\ref{fig:FCC_flu}. The solid curve in Fig.~\ref{fig:FCC_sp} was
obtained with x-ray reflection switched off. Instead, the dashed and
dotted curves were obtained with x-ray reflection activated, and an
$R_q$ parameter (see Eq.~\eqref{eq:fNC}) set to 25~nm and 0~nm
respectively, the latter implying a perfect mirror surface. X-ray
reflection at the beam pipe, collimators, and SR masks leads to an
increase by a factor of up to~60\% in photon fluence around the IP
for perfect-mirror surfaces (dotted line). This
enhancement naturally reduces with increasing surface roughness (see
the dashed curve, where a sizeable effect is
nevertheless still visible).

\begin{figure}
  \centering
  \includegraphics[width=\linewidth]{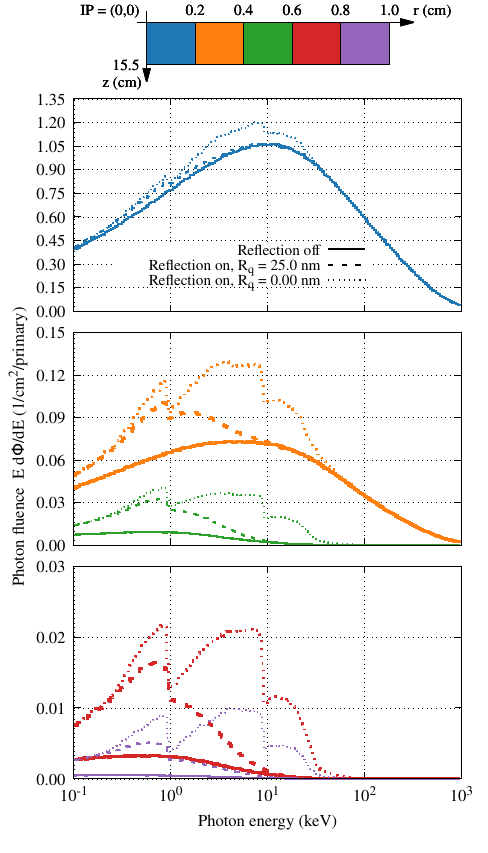}
  \caption{Energy-differential photon fluence at various radial
    distances from the IP (see colour code in top diagram and the text
    for more details), obtained with x-ray reflection deactivated (solid
    line) and activated on mirror-like surfaces (dotted line) and with a
    roughness characterised by $R_q=25$~nm (dashed line).}
  \label{fig:FCC_sp_resolved}
\end{figure}

Figure~\ref{fig:FCC_sp_resolved} further resolves the
energy-differential photon fluence in off-beam (radial) coordinate. The
top panel restricts the cylindrical scoring volume around the IP to
$r\in[0,0.2]$~cm; the middle panel to $r\in[0.2,0.4]$~cm (thicker lines)
and $r\in[0.4,0.6] $~cm (thinner lines); and the bottom panel to
$r\in[0.6,0.8]$~cm (thicker lines) and $r\in[0.8,1.0]$~cm (thinner
lines). Solid, dashed, and dotted lines carry the same meaning as in the
foregoing Figure. As anticipated above, the largest discrepancies
introduced by x-ray reflection are found at the farther off-beam radial
regions. Still, in the closest region to the IP, x-ray reflectivity
leads to differences in photon fluence of the order of 15\%.

Thus, as was the case for LEP, x-ray reflection in upstream regions
might indeed lead to a significant increase in the photon background in
the detector region of the FCC-ee. The FLUKA x-ray reflection model
presented in this work can therefore be readily employed to assess its
magnitude in the design, \textit{e.g.}, of bespoke SR masks and
collimation systems for the FCC-ee.

\section{Summary and conclusions}
\label{sec:conclusions}

Based on atomic scattering factors from evaluated
photon data libraries, a new model for the reflection of x~rays on solid
surfaces and multilayer mirrors has been developed and implemented in FLUKA~v4-6.0. The
reflectivity is evaluated as a function of the photon energy, its
incidence angle with respect to the surface, its (linear) polarisation
state, and the roughness of the surface. Both homogeneous solids and
multilayer mirrors are modelled. The reflectivity evaluated with this
new model has been benchmarked against state-of-the-art codes for x-ray
optics (XOP, XOPPY, IMD), obtaining very good agreement, both for
mirrors and multilayers.

These enhanced capabilities allow a single FLUKA run to simulate
the emission of synchrotron light by charged particles
traversing bending magnets and wigglers (feature available since release~4-3.0 of FLUKA~\cite{FLUKA1}), the transport of synchrotron
radiation through complex geometries, the reflection on multilayer
mirrors (or homogeneous solids such as the beam pipe itself), and the
further transport of reflected photons.
This integrated workflow drastically simplifies the concatenation of
tools often required in synchrotron-light beamline studies.

Two applications have been presented, showcasing the performance of this
new model. On the one hand, the reflection of synchrotron light on a
multilayer mirror in the MINERVA beamline at the ALBA synchrotron has been
explained, allowing the radiation study to be handled in a single FLUKA run, instead of
the various simulation tools and steps that were previously required. On the
other hand, we show how this new FLUKA capability allows one to assess the
photon flux throughout the machine and especially around the interaction point
of the FCC-ee due to upstream x-ray reflection within the beam pipe. Thus, the
model presented here offers enhanced capabilities in the design of accelerator
beamline elements. It is included in the public release of FLUKA as of
version~4-6.0.

\vspace{2em}

\section{Acknowledgements}

This work was done on behalf of the FLUKA.CERN Collaboration. The
activity at Argonne National Laboratory was supported by U.S. Department
of Energy, Office of Basic Energy Sciences, under contract
No.~DE-AC02-06CH11357.

\vfill

\bibliography{references}

\end{document}